\documentclass[12pt]{iopart}

\usepackage{graphicx} 
\graphicspath{{figs/}}
\usepackage{iopams}
\usepackage{color}
\usepackage{indentfirst}
\usepackage{enumerate}
\usepackage[colorlinks,
            linkcolor=blue,
            anchorcolor=blue,
            citecolor=green
            ]{hyperref}
\usepackage{cite}
\usepackage[justification=centering]{caption}

\usepackage[normalem]{ulem}

\newcommand{\tn}{\tilde{n}}

\newcommand{\eqmatrix}[1]{\left(\eqalign{#1}\right)}
\newcommand{\eqcase}[1]{\left\{\eqalign{#1}\right.}
\newcommand{\TRC}{MOE Key Laboratory of TianQin Mission, %
TianQin Research Center for Gravitational Physics $\&$ %
School of Physics and Astronomy, %
Frontiers Science Center for TianQin, %
Gravitational Wave Research Center of CNSA, %
Sun Yat-sen University (Zhuhai Campus), %
Zhuhai 519082,  %
People's Republic of China}

\begin{document}

\title{Adaptive Modeling of Correlated Noise in Space-Based Gravitational Wave Detectors}

\author{Ya-Nan Li, Yi-Ming Hu, En-Kun Li\footnote{Corresponding Author}} 
\ead{lienk@mail.sysu.edu.cn} 
\address{\TRC}
\begin{indented}
\item[] \today
\end{indented}

\begin{abstract}
    Accurately estimating the statistical properties of noise is important in data analysis for space-based gravitational wave detectors. 
    Noise in different time-delay interferometry channels correlates with each other. 
    Many studies often assume uncorrelated noise and ignore the off-diagonal elements in the noise covariance matrix. 
    This could lead to some bias in the parameter estimation of gravitational wave signals. 
    In this paper, we present a framework for reconstructing the full noise covariance matrix, including frequency-dependent auto- and cross-correlated power spectral densities, without assuming the parametric analytic expressions of the noise model. 
    Our approach combines spline interpolation with trigonometric basis functions to construct a semi-analytical representation of the noise. We then employ trans-dimensional Bayesian inference to fit the correlated noise structure.
    The resulting software package, \texttt{NOISAR}, successfully recovers both auto- and cross-correlated power spectral features with a relative error of about $10\%$.
\end{abstract}

\section{Introduction}

Ground-based gravitational wave (GW) observatories, such as LIGO\cite{Sigg_2006}, Virgo\cite{VIRGO_Collaboration:_2003}, KAGRA\cite{Somiya_2012}, and the upcoming space-based detectors, such as TianQin \cite{Mei2021, 2015TianQin}, 
Laser Interferometer Space Antenna (LISA)\cite{Karsten_Danzmann_1996}, 
and Taiji\cite{Cyranoski2016ChineseGH}, 
are designed to detect extremely weak signals from astrophysical and cosmological sources \cite{Li:2024rnk}. 
These signals are often buried in instrumental noise originating from test-mass (TM) acceleration noise, optical metrology system (OMS) noise, and laser frequency noise \cite{Babak:2021mhe, Bender:2005ue, 2015TianQin}. Although laser frequency noise can be suppressed by applying the time-delay interferometry (TDI) method \cite{Armstrong_1999, Tinto:2002de, Estabrook:2000ef}, the other two kinds remain in the data and will influence the parameter estimation precision of GW signals.
Accurate characterizing and modeling of the noise are critical for distinguishing between noise and genuine GW signals, especially for stochastic gravitational wave backgrounds (SGWBs)\cite{Regimbau:2011rp, Maggiore:2018sht, Christensen_2019, Cheng:2022vct}.

Space-based detectors are designed to observe GWs from various millihertz-band sources, e.g., ultracompact binaries (UCBs), binary black holes (BBHs), et al. These signals help constrain the origin and evolution of compact binaries and the structure of the Milky Way \cite{Korol:2018wep, Adams:2012qw}. 
Sensitive detection and accurate parameter estimation of the GW signals are thus important for us when exploring the rich information. 
A fundamental challenge in space-based GW data analysis arises from the sheer number (tens of thousands) and long duration of detectable signals, which overlap in both the time and frequency domains. To simultaneously recover all resolvable signals from noisy data and avoid cumulative errors from hierarchical fitting, a global fit approach is employed, jointly modeling all sources and noise \cite{Littenberg:2023xpl, Strub:2024kbe, Katz, Deng:2025wgk}.
A critical step in this analysis is reconstructing the noise properties, for Gaussian noise shown as a noise covariance matrix. The diagonal elements are the variance of each parameter, and the off-diagonal ones are the covariance between different parameters.


To suppress laser frequency noise, which is much stronger than GW signals, the TDI method is employed in data processing \cite{Armstrong_1999, Tinto:2002de, Estabrook:2000ef}. 
Michelson combinations of TDI channels, e.g., $X$, $Y$, $Z$, contain highly correlated noises.
Quasi-uncorrelated TDI channels $A$, $E$, and $T$ are orthogonal (or uncorrelated) under the assumptions of equal arm lengths and identical instrument noise. 
To simplify complex parameter estimation tasks, most studies consider utilizing quasi-orthogonal $A$, $E$, and $T$ channel data by assuming a diagonal noise covariance matrix. 
However, in practical scenarios, the arm lengths and instrument noise are rarely identical, which means that the off-diagonal elements of the noise covariance matrix cannot be neglected. 
In global analysis pipelines, where signals from various sources and noise are alternately updated, a more complete noise model can significantly improve the accuracy of parameter estimation\cite{Littenberg:2023xpl, Strub:2024kbe, Katz, Deng:2025wgk}. 
Therefore, to avoid bias in GW signal parameter estimation, it is essential to estimate the full noise covariance matrix.


So far, some researchers in the field of SGWB have taken a more comprehensive approach to noise modeling, since correlated noise can mimic or obscure SGWB signals and significantly affect detection capabilities \cite{Cheng:2022vct}. 
Adams et al. \cite{Adams:2010vc} distinguish between SGWB and instrumental noise while considering a full noise covariance matrix. Each element of the matrix is derived from known analytical expressions.
Baghi et al. \cite{Baghi:2023qnq} develop a Bayesian framework and use cubic splines to model instrumental noise with an unknown spectral shape. This approach enables robust SGWB detection without prior assumptions on noise characteristics.
They ignore the laser noise and assume that there are no correlations between single-link measurements. 
Caporali et al. \cite{Caporali:2025mum} study the impact of correlated noise on SGWB detection and relative parameter estimation for the next generation ground-based GW interferometer, the Einstein Telescope (ET) \cite{Punturo:2010zz, ET:2019dnz}. 
They perform a Bayesian analysis on simulated data and illustrate that neglecting correlated noise introduces significant biases in the parameter reconstruction of SGWB.

Various methods have been developed for noise reconstruction. 
The typical \texttt{Welch} method \cite{Welch:1967oth} computes the median-averaged power spectrum of ``off-source'' data segments near a candidate signal. The result can be biased due to drifts in noise levels and noise transients\cite{Chatziioannou:2019zvs, Rover:2008yp, Talbot:2020auc}. 
To address these issues, ``on-source'' spectral estimation methods have been developed, such as $\mathtt{Bayesline}$ and $\mathtt{Bayeswave}$. 
These methods combine cubic splines for broadband noise and different models for narrowband noise. They take into account the multi-component and variable dimension, leveraging the advantage of the trans-dimensional Bayesian approach\cite{Littenberg:2014oda, Cornish:2014kda}. 
These two methods have been shown to effectively recover auto-correlated noise power spectra for both ground-based and space-based gravitational wave detectors\cite{Cornish:2020dwh, Becsy:2016ofp, Gupta:2023jrn}. Besides, some noise estimation research employs theoretical parametric expressions for the noise model and estimates the associated noise parameters \cite{PhysRevD.111.064056}.


In this paper, we aim to develop a software called \texttt{NOISAR} (NOIse Space Adaptive Reconstruction) to reconstruct the full noise covariance matrix, including correlated components. Our method combines spline interpolation with a trigonometric function to semi-analytically fit both the Power Spectral Density (PSD) and the Cross Spectral Density (CSD) between different channels. This approach is phenomenological. 
Users do not need to know the analytical form of the noise model in advance. 
They only need to provide the noise data, and the software will return the noise covariance matrix after processing. 

\texttt{NOISAR} can be integrated into parameter estimation workflows for various signals, such as global analysis. 
The global analysis pipeline for space-based detectors will simultaneously recover signal parameters from multiple sources and noise components. With \texttt{NOISAR} as the noise fitting module, the pipeline updates the noise PSD and CSD in each iteration and thus enhances the performance of parameter estimation of GW signals.
It is important to note that the current version of the software, developed in this study, is preliminary. It is designed to model pure Gaussian and stationary instrumental noise. 
However, the method shows potential for handling foreground noise from unresolvable sources. Since it is phenomenological, it can recover the shape of the PSD curve resulting from the combined effects of foreground and instrumental noise. 
This work does not yet address more complex scenarios, such as data gaps, transient noise in the data stream, and non-stationary noise. These challenges will be incorporated into the advanced \texttt{NOISAR} framework in future developments. In this paper, as a simple case, we demonstrate an example of fitting the noise PSD and CSD for $X, Y, Z$ channels with Gaussian and stationary instrument noise. 

The outline of this paper is as follows: Sec.~\ref{sec:noise_theory} introduces the theoretical noise power spectra of TianQin detectors in the $X$, $Y$, and $Z$ channels. Sec.~\ref{sec:methodology} provides a detailed description of the methods used in this study, including noise modeling and the Bayesian inference approach. Sec.~\ref{sec:result} presents the fitting results for the PSD and CSD, comparing them with theoretical ones. Sec.~\ref{sec:conclusion} concludes the paper and discusses the scope of this method’s application and its future development.

\section{Instrument Noise In Detector Channels}\label{sec:noise_theory}

Different TDI channels are defined as linear combinations of time-shifted single-link components. In this study, we focus on the $X$, $Y$, and $Z$ TDI channels, which are correlated with each other whenever the arm lengths are equal. 

Assuming that the three spacecrafts are all identical, the analytical formula of PSD and CSD for $X$, $Y$, and $Z$ channels can be written as\cite{Armstrong_1999}:
\begin{eqnarray}
    S^n_X &=& 16 \sin^2 u \left[ S_s + 2(1+\cos^2 u) S_a \right], \\
    S^n_{XY} &=& -8\sin^2 u \cos u \left[ S_s + 4 S_a \right],
\end{eqnarray}
where $u=2\pi fL/c$, $c$ is the speed of light, and $L\approx 1.7\times 10^8\, \rm m$ is the arm length of TianQin.
$S_a$ and $S_s$ are acceleration and readout noise (mainly shot noise) for TianQin, which are given by:
\begin{eqnarray}
    S_{a}(f) = \frac{N_a}{(2\pi f c)^2}
    \left(1+\frac{0.1 \rm mHz}{f} \right),
    \quad 
    S_{s}(f)=N_s \left(\frac{2\pi f}{c}\right)^2,
\end{eqnarray}
where $\sqrt{N_a} =3\times 10^{-15} {\rm m/s}^{2} {\rm Hz}^{-1/2}$ and $\sqrt{N_s} = 1 {\rm pm \, Hz}^{-1/2}$  are the designed requirement for acceleration and readout noise \cite{2015TianQin}. 

\begin{figure}[ht]
    \centering
    \includegraphics[width=0.7\linewidth]{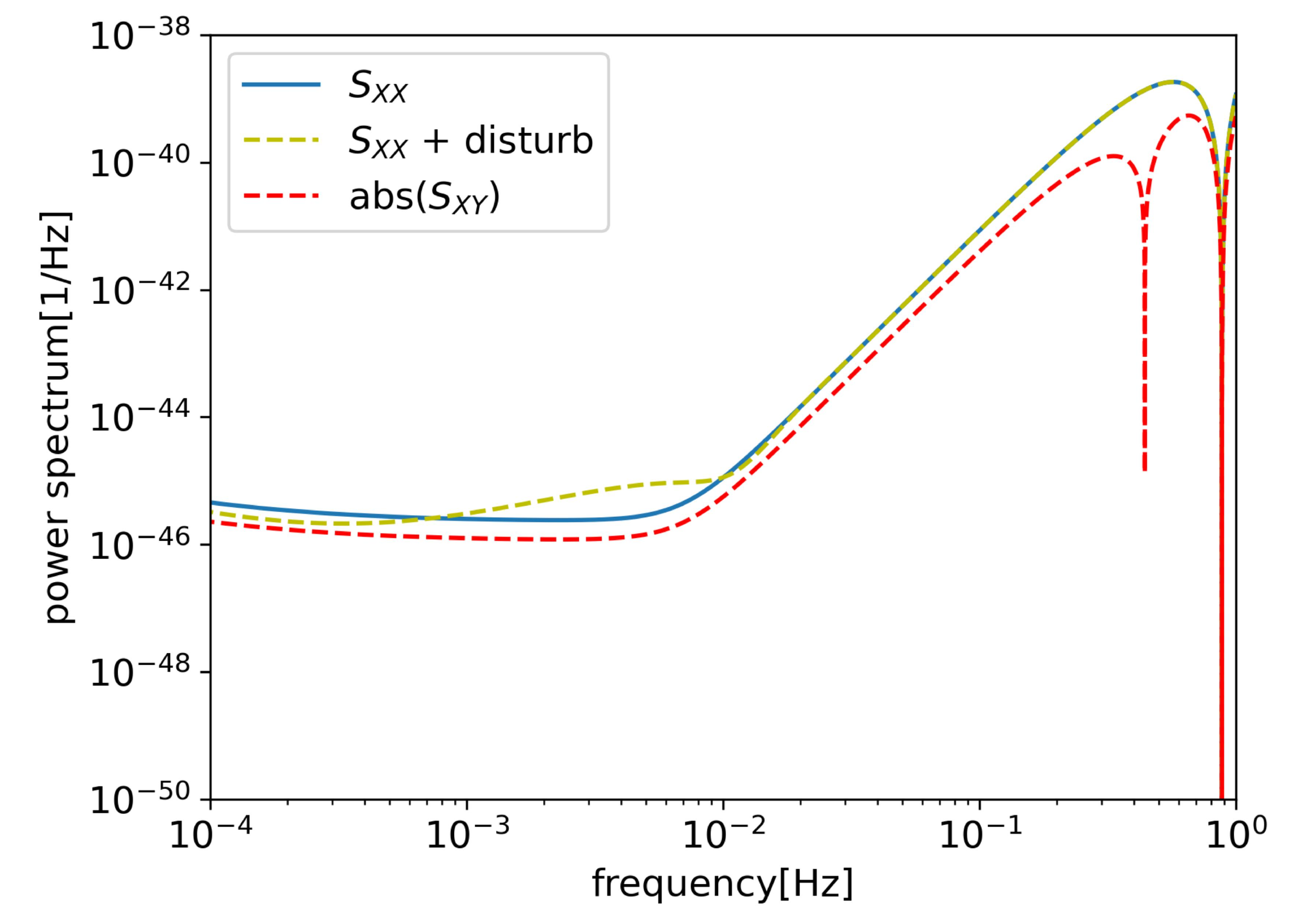}
    \caption{Analytical PSD (blue solid) and absolute value of CSD (red dashed). Disturbed PSD is shown as a yellow dashed line.}
    \label{fig:analy}
\end{figure}

FIG.~\ref{fig:analy} shows the PSD and the absolute value of CSD in logarithmic coordinates as a function of frequency. 
To simulate more realistic noise, whose PSD curve is not ideally flat in the low-frequency region, we introduce a small disturbance on top of the theoretical PSD. The average fluctuation across the frequency range is $\sim 1\%$ of the theoretical PSD, and the value of the maximum fluctuation point is about three times the theoretical PSD.

Notice that there is a descending peak in the PSD curve and two peaks in the CSD curve. 
The peak at the higher frequency, which appears in both the PSD and the CSD, results from the spectrum values that transition from positive to zero. 
It's mathematically due to the $\sin^2(f/f^*)$ in detector response functions, where $f^*\equiv c/2\pi L$ is the ``transfer frequency''. 
The other peak arises from the CSD crossing the zero point, transitioning from negative to positive values, due to the $\cos (f/f^*)$ term. We also aim to fit these important features accurately.

In this paper, we use the disturbed PSD and analytical CSD shown in FIG.~\ref{fig:analy} to generate correlated noise data. Therefore, in this simple case, we follow the assumption that $S_{XX}=S_{YY}=S_{ZZ}$ and $S_{XY}=S_{XZ}=S_{YZ}$. 
However, this assumption is not required and our method can be applied to more general cases.
The details of data generation are provided in the following section.


\section{Methodology}
\label{sec:methodology}

\subsection{Preprocessing Phase}\label{sec:preprocess}

We generate correlated noise data $n(f)$ using the frequency-dependent noise covariance matrix $\mathbf{C}(f)$. The diagonal elements of $\mathbf{C}(f)$ represent the auto-PSDs of $X$, $Y$, while $Z$ channels, and the off-diagonal elements represent the CSDs between two different channels. 
In real scenarios, if the detected data is in the time domain, one should first Fourier transform it into the frequency domain, since the PSD and CSD estimation is performed in the frequency domain in this method.

In this section, we aim to smooth the raw data and get an initially averaged spectrum according to the definition of covariance:
\begin{eqnarray}
    {\rm Cov}(X(t),Y(t)) &= \left\langle (X(t)-\langle X(t) \rangle)(Y(t)-\langle Y(t) \rangle) \right\rangle, 
\end{eqnarray}
where $X(t)$ and $Y(t)$ represent the time domain data in channel $X$ and $Y$, $X(t_i)$ and $Y(t_i)$ are the $i$th points in $X(t)$ and $Y(t)$. $n$ is the total number of averaged data points. $\langle \cdot \rangle$ denotes the expected value.

In the frequency domain, the variance and covariance become:
\begin{eqnarray}\label{eq:psd0}
    {\rm PSD} &=& 2\left\langle \tn_X(f)\tn_X^*(f') \right\rangle / \delta(f-f'), \\\label{eq:csd0}
    {\rm CSD} &=& 2\left\langle \tn_X(f)\tn_Y^*(f') \right\rangle / \delta(f-f'),
\end{eqnarray}
where $\tn_X(f)$ and $\tn_Y(f)$ are frequency domain data after the Fourier transform, and the asterisk denotes the complex conjugate. $\delta$ represents the Dirac delta function.
The expected value is calculated by averaging over $n$ data points: 
\begin{eqnarray}\label{eq:ave1}
\left\langle \tn_X(f)\tn_X^*(f') \right\rangle = \frac{1}{n} \sum_{i=1}^{n}\left( \tn_X(f)\tn_X^*(f')\right), \\
\label{eq:ave2}
\left\langle \tn_X(f)\tn_Y^*(f') \right\rangle = \frac{1}{n} \sum_{i=1}^{n}\left( \tn_X(f)\tn_Y^*(f') \right).
\end{eqnarray}

In reality, instead of directly averaging by Eq.~(\ref{eq:ave1}) and~(\ref{eq:ave2}), we take the median of every $n$ points to avoid the influence of outliers that deviate significantly from the mean value. Then we obtain initial median smooth spectra $D_{XX}$ and $D_{XY}$.
Notice that the product of the noise data in different channels (e.g., $\tn_X(f)\, \tn_Y^*(f')$) is complex. The imaginary part oscillates around zero, making it difficult to fit a smooth curve phenomenologically. Therefore, in this work, we fit the real part to recover the power of the cross-correlated noise spectrum. Since the magnitudes of the real and imaginary parts are comparable, the real part differs from the complex quantity by a factor of $\sqrt{2}$.

Using the definition of PSD and CSD in Eq.~(\ref{eq:psd0}) and~(\ref{eq:csd0}), we obtain the spectra directly from the noise data.
However, strong random fluctuations prevent them from serving as a universal representation of the features (see the light blue curves in FIG.~\ref{fig:spline_ini}). It is necessary to smooth it out to serve as a noise model. 
The noise modeling method described in the following sections is based on those median smooth spectra. 

\subsection{Modeling the noise covariance matrix}
\label{sec:model}

\subsubsection{Spline Parameterization}
\label{sec:spline}

In theory, expressing an unknown function as an $N$-degree polynomial allows for the derivation of the functional relationship between the independent and dependent variables by obtaining its coefficient matrix. Here, the polynomial can be written as:
\begin{eqnarray}
    S &=& \sum_{i=0}^N a_i (f - f_i)^i,
\end{eqnarray}
where $S$ represents the fitted model, $N$ is the degree of the polynomial expansion, and $a_i$ is the coefficient to be estimated. 
Calculating an $N$-degree matrix is time-consuming. In practice, we use the spline interpolation method to model the entire curve with only tens or hundreds of knots. To avoid large oscillations between knots in cubic spline curves, we use the B-spline-based \texttt{Python} interpolation package $make\_interp\_spline$. This method ensures that the interpolated curve passes through each knot. 
Due to the noise data spanning multiple orders of magnitude, direct interpolation yields poor results. Therefore, we first apply a logarithmic transformation to the data, allowing for smoother interpolation within the same order of magnitude. Afterward, the data is converted back to the original scale. Since the CSD contains negative values, taking the logarithm directly is not feasible. Instead, we use a segmented logarithmic transformation method: we first take the logarithm of the absolute values and then restore the signs. A detailed description of this method is provided in Sec. \ref{sec:log}.

The initial knots are selected from the log-transformed median smooth spectra PSD$_0$ and CSD$_0$ obtained in previous steps. 
Here we choose about 40 knots log-uniformly in the frequency region below 20 mHz. The number of knots is determined empirically after several rounds of trials, to ensure a smooth curve that can roughly recover the feature of the noise spectra. 
The log-uniform choice is justified in our case because we consider pure Gaussian noise, whose spectral density exhibits a simple and smooth profile in the low-frequency region. In more complex scenarios, e.g., when there are non-stationary noise, glitches, or data gaps in the data stream, a more data-driven knot selection approach may be required. For example, one could average the data over different orders (e.g., averaging every 32 points and over 64 points) to obtain two average-smoothed curves. Knots should then be selected at locations where the difference between the two curves exceeds a certain threshold.
Notice that in the high-frequency region (where $f>20$ mHz), the PSD and CSD exhibit a drop to zero, mathematically due to the $\sin^2(f/f^*)$ term. Besides, the CSD passes through zero at a certain point due to the $\cos (f/f^*)$ term. To recover those features, we select more knots linearly near the zeros. Specifically, we set about 10 knots in the region surrounding each zero point. The width of each region is $\sim$0.05Hz. The number of knots is also determined empirically. In the subsequent Bayesian analysis, the number of knots will be updated by applying the trans-dimensional method.

\begin{figure}[htbp]
    \centering
    \includegraphics[width=0.8\linewidth]{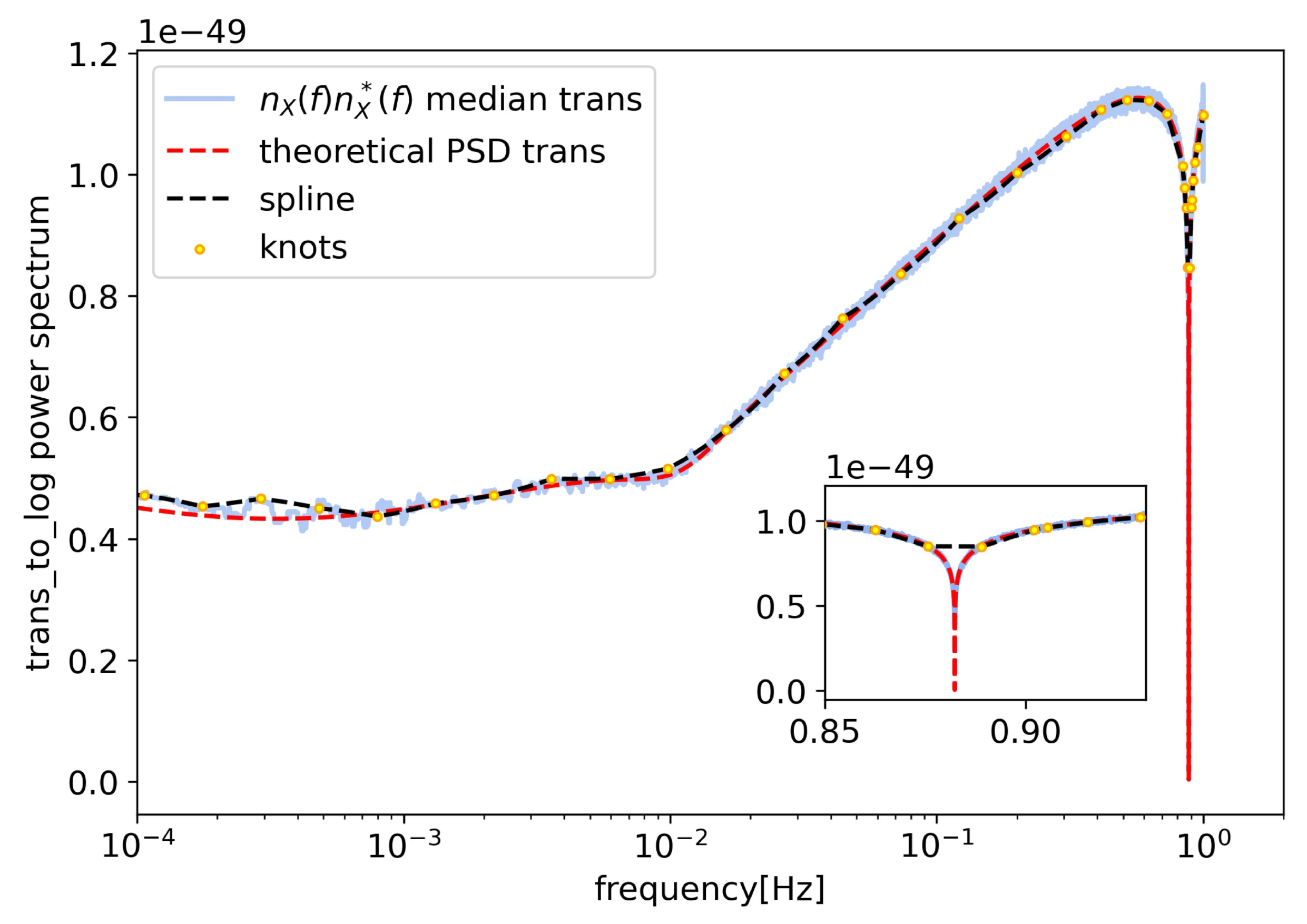}
    \includegraphics[width=0.8\linewidth]{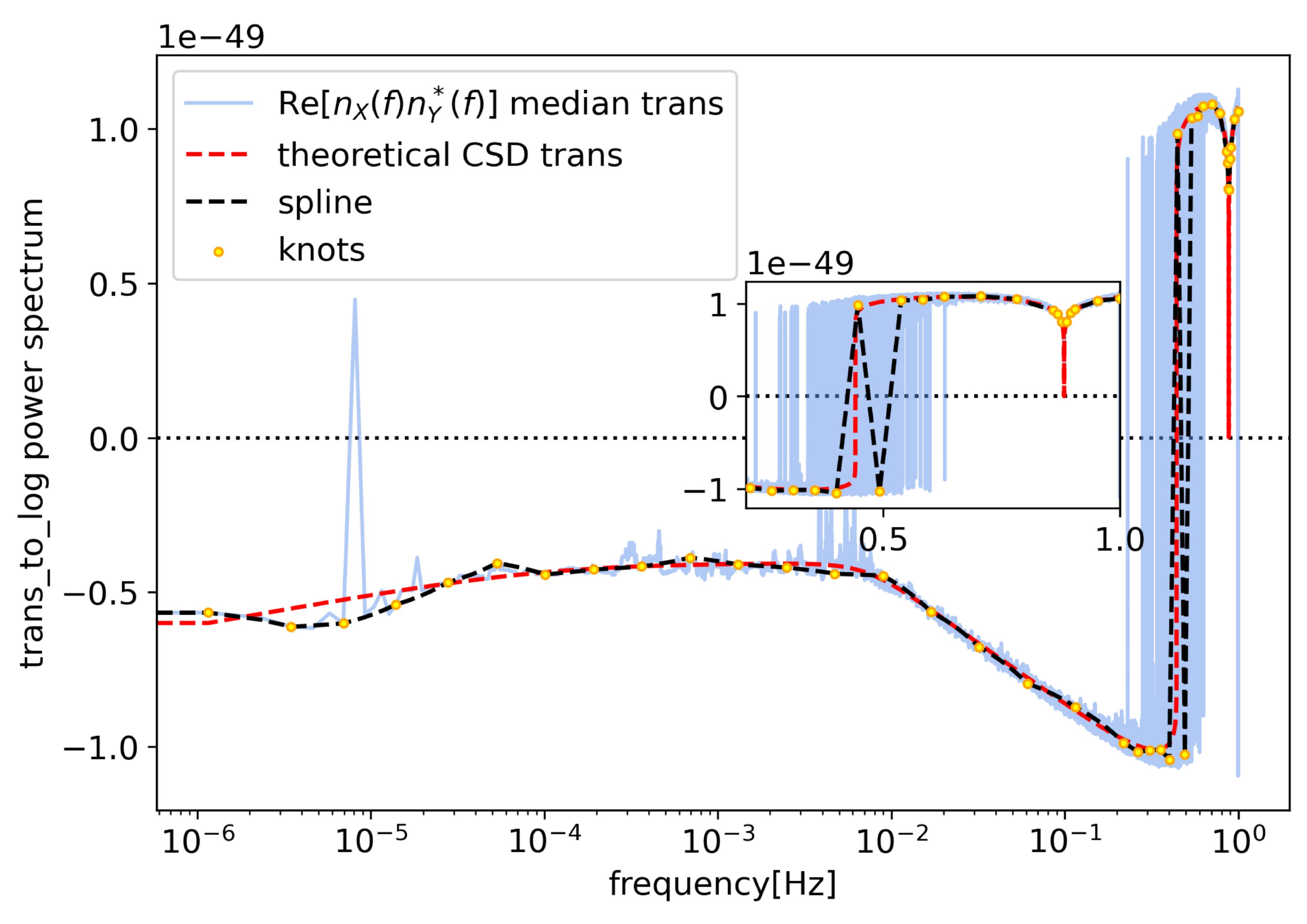}
    \caption{Median estimate for the spectrum after log-transforming (blue) and the initial spline (dashed black) on a logarithmic frequency x-axis, and with a subplot on a linear x-axis. The subplot shows a zoomed-in view of the high-frequency region. The yellow dots are knots picked from the median data spectrum, and the interpolation curve passes through these points. The upper panel shows results from single-channel data, and the bottom panel shows results from correlated data.}
    \label{fig:spline_ini}
\end{figure}

FIG.~\ref{fig:spline_ini} shows the initial log-transformed median smooth spectra with light blue lines, knots with yellow dots, and the spline curve with black lines. 
To observe the interpolation results near the zeros, we show a zoomed-in view of the high-frequency region in a subplot, displayed on a linear horizontal axis. 
The figure shows that the spline interpolation does not capture the characteristics where the curve drops to zero. 
Additionally, since there are many cross-zero points in the correlated data, spline interpolation may result in multiple cross-zero features in the recovered CSD curve, which differs from the theoretically expected result that has only one. 
To address this issue, we adopt a semi-analytic noise model, combining spline interpolation for phenomenological fitting and a trigonometric function (described in Sec. \ref{sec:trig}) to reproduce the sharp drop features in the high-frequency region.

\subsubsection{Trigonometric Function}\label{sec:trig}

As mentioned before, we use trigonometric functions to recover the power spectrum near zero points:
\begin{eqnarray}
    \label{eq:hpsd}
    h_{PSD}(f) &=& A_{PSD}(f) \sin^2(2 \pi f/f^*),\\
    \label{eq:hcsd}
    h_{CSD}(f) &=& A_{CSD}(f) \sin^2(2 \pi f/f^*) \cos(2 \pi f/f^*),
\end{eqnarray}
where $A(f)$ is a polynomial function of $f$ with multiple power terms. Empirically, a cubic term ($f^3$) as the highest order is sufficient for fitting the TianQin PSD and CSD, i.e., $A(f)=af^3+bf^2+cf+d$. The coefficients $a, b, c, d$ are the parameters we aim to fit in the following Bayesian analysis process. With this highest order, the function can roughly recover the shape of the noise spectra in the high-frequency region. The zero-crossing point is caused by the $\cos(2 \pi f/f^*)$ term, thus, this term only appears in Eq.~(\ref{eq:hcsd}) for CSD fitting. Note that here we do not vary the coefficients in the $\sin$ and $\cos$ terms. This implies an underlying assumption that the detector's three arms are of equal length, which holds for the first-generation TDI. 
In the second-generation TDI scenario, the coefficients of those two terms require subsequent fitting through Bayesian analysis.

The frequency band where the spline interpolation curve is replaced by a trigonometric function depends on the location of the zero points in the data.
Theoretical CSD of TianQin has only one cross-zero point below 1 Hz, but the generated data and the initial smoothed spectrum $D_{XY}$ have a lot of such points, see the light blue line in the right panel of FIG.~\ref{fig:spline_ini}. 
Direct interpolation would lead to more than one cross-zero point in the CSD curve, which is not what we expect. 
Thus, the frequency band covering all those points is where we adopt the trigonometric function to fit the CSD. 
Additionally, another sub-band, covering the second zero point that appears in both the PSD and CSD, is set with a width of $\sim 4$ mHz. 
This is an entirely empirical choice, using the trigonometric function to fit the peak only within a narrow range near the zero point.

We combine the B-spline interpolation and the trigonometric function to fit the whole spectrum curve. An important issue is ensuring a smooth transition boundary between the two parts. In this study, the criterion for determining smoothness is whether the relative difference between the spline interpolation value ($S_{spline}$) and the trigonometric function value ($S_{trig}$) at the connection point is smaller than a predefined threshold. The relative difference is calculated by $|(S_{trig}-S_{spline})/S_{spline}|$. The ``connection point'' refers to the frequency point that serves as the boundary between the two modeling methods. In this paper, the threshold is artificially set to 20$\%$. 
In each iteration of the following Bayesian analysis process, if the relative difference exceeds the threshold, the proposal point will be rejected. Additionally, the initial coefficient values $a_{ini}, b_{ini}, c_{ini}, d_{ini}$ are also chosen according to this criterion. A lower threshold can improve smoothness at the connection point, but may reduce the acceptance rate in subsequent Bayesian updates.

\subsection{Logarithmic method of the data}
\label{sec:log}

This section aims to transform the initially average smoothed spectrum into logarithmic form, allowing for spline interpolation within the same order of magnitude and resulting in smooth curves.

For correlated noise data, the CSD contains both positive and negative values. 
Thus, directly taking the logarithm is infeasible.
To address this issue, we segment the spectrum into three parts based on its sign and magnitude: positive part (named ``+''), negative part (named ``$-$''), and a transition part near zero. For the points with values within a small range around 0 (with $S_{\rm th}$ as the boundary, $S_{\rm th}$ is chosen artificially and will be described below), retain their original sizes. 
For data points with absolute values greater than $S_{\rm th}$, we take the logarithm of the positive part directly. The negative part is first converted to its absolute value, then the logarithm is taken, and finally the negative sign is restored. The conversion formula is:
\begin{eqnarray}\label{eq:log}
    S_{i\_log} &=& \eqcase{
    S_{\rm th} \log_{10} (S_i/S_{\rm th}), & {\rm ~ for~ } S_i > S_{\rm th}, \\
    S_i, & {\rm ~ for~ } - S_{\rm th} \leq S_i \leq S_{\rm th}, \,    (i=1,\ldots,N)\\
    -S_{\rm th} \log_{10} (-S_i/S_{\rm th}), & {\rm ~ for ~} S_i < -S_{\rm th},
    }
\end{eqnarray}
where $S_i$ is the value of each point in the spectrum we want to transform, and $S_{i\_log}$ is the spectrum after being logarithmized. $N$ is the number of data points. $S_{\rm th}$ is some threshold value which defines the range of the transition part, and is set to $S_{\rm th} = 1\times 10^{-50}$ in this paper. This value is artificially selected. Any value is valid as long as its magnitude is smaller than most of the points in the target spectra; the smaller, the safer. The reason is described in the following paragraph.

It is established that the exponents of the spectrum points $S_i$ with a base of 10 are negative. Here, for example, one of the points $S_i=10^{-30}$ is a positive value in the ``+'' part. After taking the logarithm directly (also based on 10), it comes to $\log_{10}(S_i)=-30$, which is a negative value. Obviously, the signs of each point after transformation are opposite to what was expected. 
A natural consideration is then simply multiplying them by $-1$ and reversing the signs. 
However, it is invalid because the relative magnitudes of the values are also inverted. To address this issue, before taking the logarithm of the data, we divide it by $S_{\rm th}$, whose exponent is smaller than those of the ``+'' and ``-'' segments. Here, as an example, $S_i/S_{\rm th}=10^{-30}/10^{-50}=10^{20}$. This division transforms the exponents into positive numbers, without affecting the relative magnitudes of the original values. 
Finally, we multiply the logarithmic values by $S_{\rm th}$, bringing all segments' magnitudes close to $S_{\rm th}=10^{-50}$, which allows for smooth spline interpolation. Note that the transition part between the positive and negative part, shown as the second line of Eq.~(\ref{eq:log}), does not need any operation. 
Because its magnitude is so small and close to zero, it has minimal impact on smooth interpolation.

Mathematically, $S_{\rm th}$ can be set to any value artificially. However, if it is so large that most of the points' magnitude is smaller than the threshold, the transition part will cover almost the whole frequency region, and no transformation will be done. 
Therefore, $S_{\rm th}$ should be a very small value, as discussed, smaller than most of the points in the target CSD. 

\subsection{Bayesian analysis}

\subsubsection{Likelihood}
The spline knots and trigonometric function coefficients are updated utilizing the Markov Chain Monte Carlo (\texttt{MCMC}) sampling method \cite{Hastings1970} based on Bayesian inference to fit the target PSD and CSD. 
Crucially, the number of spline knots is dynamically updated via the trans-dimensional reversible jump \texttt{MCMC} (\texttt{RJMCMC}) method\cite{RJMCMC}, allowing the program to adaptively fit the curves that best describe the noise data. 
This approach is similar to the $\mathtt{Bayeswave}$ method, while this paper considers the full noise covariance matrix when constructing the likelihood function:
\begin{equation}
    \mathcal{L}(\mathbf{d}|\boldsymbol{\theta}) = \prod_f \frac{1}{\sqrt{(2\pi)^N |\mathbf{C}(f, \boldsymbol{\theta})|}} \exp\left(-\frac{1}{2} \mathbf{d}^\dagger(f) \mathbf{C}^{-1}(f, \boldsymbol{\theta}) \mathbf{d}(f)\right),
\end{equation}
where $\boldsymbol{\theta}$ includes spline parameters and trigonometric function coefficients, $\mathbf{C}$ is the noise covariance matrix, and $\mathbf{d}$ is residual data after subtracting GW signals and cleaning glitches, ideally only contains Gaussian stationary noise. Taking the logarithmic form and expanding the vectors, we have: 
\begin{eqnarray}\label{eq:like}
    \log \mathcal{L} &=&
    - \frac{1}{2}
        \eqmatrix{n_X^*(f) \; n_Y^*(f) \; n_Z^*(f)}
    \mathbf{C}^{-1}
    \eqmatrix{n_X(f) \\ n_Y(f) \\ n_Z(f)}
    \nonumber \\ 
    && - \frac{1}{2} \log ((2\pi)^N \det\mathbf{C}),
\end{eqnarray}
where
\begin{equation} \label{eq:cov}
    \mathbf{C} = \eqmatrix{
        S_{XX}(f) \;  S_{XY}(f) \; S_{XZ}(f)\\
        S_{XY}^*(f) \; S_{YY}(f) \; S_{YZ}(f)\\
        S_{XZ}^*(f) \; S_{YZ}^*(f) \; S_{ZZ}(f)
    }.
\end{equation}
The diagonal elements are the PSDs, while the off-diagonal elements are the CSDs we need to fit.

The prior distribution imposes upper and lower bounds on the number of knots and enforces relative smoothness at some connection points between spline interpolation and trigonometric function fitting.

\subsubsection{Process}

Theoretically, PSDs and CSDs are updated simultaneously during the \texttt{RJMCMC} process. After sufficient iterations, the sampling chain will converge to a stationary posterior distribution. However, due to the combined influence of PSDs and CSDs on the computed likelihood value, there may be situations where the two curves deviate significantly from theoretical expectations individually, yet the resulting posterior probability value is accepted. This reduces the efficiency of the fitting and increases the number of iterations required to achieve convergence. To address this problem, referencing the \texttt{Gibbs} sampling methodology\cite{Gibbs1, Gibbs2}, we decouple the fitting of each element in the matrix $\mathbf{C}$. The main idea of \texttt{Gibbs} sampling is that, assuming the parameters can be grouped into distinct subsets $\Theta_1, \Theta_2, \Theta_3, ...$, when updating a set of parameters $\Theta_1$, the other sets of parameters $\Theta_2, \Theta_3, ...$ are kept fixed. After completing the update for this iteration, $\Theta_1$ is fixed at its current position, and the process moves on to update $\Theta_2$, and so on. Here, we illustrate a simplified scenario where $S_{XX}=S_{YY}=S_{ZZ}$, $S_{XY}=S_{XZ}=S_{YZ}$, implying that only one $S_{XX}$ and one $S_{XY}$ in the matrix require fitting.

\begin{figure}[htbp]
    \centering
    \includegraphics[width=1\linewidth]{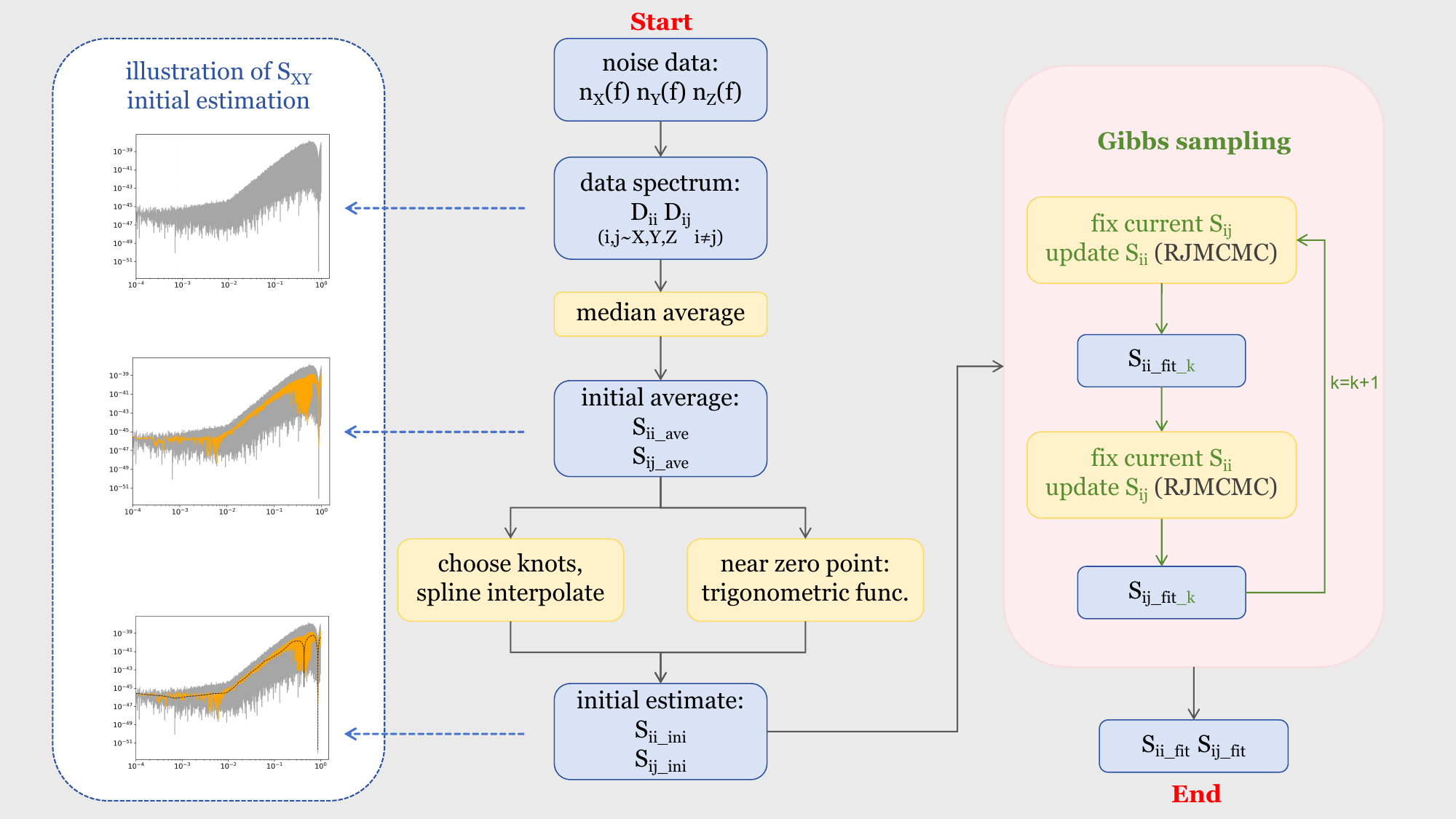}
    \caption{Flow diagram of the whole process of \texttt{NOISAR}. The process on the left side shows the initialized PSD and CSD, while the right side involves the trans-dimensional updating of PSD and CSD within the Gibbs framework.}
    \label{fig:NOISARprocess}
\end{figure}

The workflow of our method, shown in FIG. \ref{fig:NOISARprocess}, is as follows:

\begin{enumerate}[1)]
\item As described in Sec. \ref{sec:preprocess}, \ref{sec:model}, obtain the median smoothed spectra from the data and select knots on it to perform spline interpolation. Near the two descent peaks in the high-frequency region, replace the interpolation curve with a trigonometric analytic expression. One obtains initial smooth curves $S_{XX{\rm ini}}$, $S_{XY{\rm ini}}$.
\item Fix CSD at the current state (e.g., in the first iteration it is $S_{XY{\rm ini}}$), independently fit PSD. Update the analytical expression coefficients, and perform a trans-dimensional update of the spline knots followed by re-interpolation. $S_{XX{\rm fit\_k}}$ is obtained in the $k$th Gibbs iteration.
\item Fix PSD at $S_{XX{\rm fit\_k}}$, independently fit CSD this time, and get $S_{XY{\rm fit\_k}}$.
\item Repeat steps 2 and 3 iteratively, and the final fitting results $S_{XX{\rm fit}}$, $S_{XY{\rm fit}}$ are obtained.
\end{enumerate}

\subsubsection{An Acceptance Issue}

In the \texttt{RJMCMC} process, a crucial aspect is the criterion used to determine whether to accept the proposed points at each iteration. 
The typical acceptance ratio is calculated by:

\begin{equation}
\alpha\left(\theta_{i}, \theta^{*}\right) \equiv \min \left(1, \frac{P\left(\theta^{*} \mid D\right)}{P\left(\theta_{i} \mid D\right)} \frac{q\left(\theta^{*} \mid \theta_{i}\right)}{q\left(\theta_{i} \mid \theta^{*}\right)}\right)
\end{equation}
where $\theta_i$, $\theta^*$ are the parameters of current state and proposed state, $P$ is the posterior distribution and $q\left(\theta^{*} \mid \theta_{i}\right)$ is the proposal density from current state $i$ to proposed state. According to this formula, the procedure tends to accept proposal points in the high posterior region. 
The chain converges when it satisfies the detailed balance condition, which means the transition probabilities between any two adjacent states satisfy microscopic reversibility.

The algorithm tends to accept states with higher posterior values, which in this study can be interpreted as favoring higher likelihood. 
However, after several runs of the program, we observed that the knots located in the ``concave'' part (see $\sim$ 10mHz) and the ``convex'' part (see $\sim$ 4mHz) of the PSD curve are often deleted during the trans-dimensional update process, particularly the concave ones. In that case, it becomes challenging to fit the fluctuating portions of the curve accurately. 

This may be due to two reasons: 

1) The log-likelihood value is obtained by summing the values of each point in the frequency domain. Therefore, some points with higher values and some points with lower values may lead to a similar result, and the program tends to accept a simpler model with fewer concave and convex fluctuations. 

2) To illustrate intuitively, the likelihood function shown as Eq.~(\ref{eq:like}) can be written in a rough form: $-\mathbf{d C^{-1} d^*} - {\rm log\_term}$. 
Since the data $\mathbf{d}$ remains unchanged, an increase in $\mathbf{C}$ leads to a larger likelihood value, which can be accepted by the program.
When fitting PSD independently in step 2 described above, a higher curve would be selected. 
Although the log\_term restricts it from increasing indefinitely, it is likely to choose a curve that is higher and has a simpler shape in the concave parts.

To address this issue, we introduce an additional criterion for determining whether to accept the update. When the difference between the proposed point and the current point is less than $3\sigma_{PSD}$, the update is accepted; otherwise, it is rejected. 
$\sigma_{PSD}$ represents the variance of the spectrum obtained by averaging over the data spectrum. The derivation process for calculating $\sigma_{PSD}$ is as follows.

The definition of PSD is:
\begin{equation}
    {\rm PSD}=\frac{1}{m} \sum_{i=j}^{j+m} ( \tn_X(f)\tn_X^*(f') )_i,
\end{equation}
where $m$ represents the length over which an average is performed. To simplify the expression, let $D{\rm sum}$ denote $\sum_{i=j}^{j+m} ( \tn_X(f)\tn_X^*(f') )_i$ we have:
\begin{equation}
\label{eq:sigma1}
    \sigma_{\rm PSD}=\frac{1}{m} \sigma_{D{\rm sum}}
\end{equation}
Since the data $n_X(f)$ follows a Gaussian distribution with PSD as variance, $\frac{D{\rm sum}}{PSD}$ should follow a $\chi^2$ distribution with $m$ degrees of freedom, whose expected value is $m$ and variance is $2m$. That is:
\begin{equation}\label{eq:sigma2}
    \sigma \left( \frac{D{\rm sum}}{\rm PSD} \right) =\sqrt{2m}
\end{equation}
In this equation, the denominator represents the theoretical PSD that the noise data follows, which is a constant. Therefore, it can be factored out of the parentheses. In practical scenarios, this PSD is unknown, so we can substitute it with the initial PSD$_0$ obtained from Eq.~(\ref{eq:psd0}). Combine Eq.~(\ref{eq:sigma1}) and Eq.~(\ref{eq:sigma2}), we have:
\begin{equation}\label{eq:sigma}
    \sigma_{\rm PSD} =\sqrt{\frac{2}{m}}{\rm PSD}_0
\end{equation}
This additional acceptance criterion aims to limit the deviation of the updated spectrum from the initial spectrum. As mentioned above, solely pursuing higher likelihoods and simpler curves would result in the loss of much of the information in the true spectrum. 
CSD fitting follows the same approach as described above.




\section{Implementation and Results}
\label{sec:result}

Taking the $X$ and $Y$ channels as an example, we generate correlated noise using the full noise covariance matrix $\mathbf{C}$, which is constructed from the theoretical TianQin PSD and CSD. 
The observation duration is 10 days, with a sampling interval of 0.5 seconds. 
Here, we assume Gaussian and stationary noise. This paper is a preliminary proof-of-concept for the method.
More complex scenarios will be addressed in future improvements of \texttt{NOISAR}, including confusion noise, glitches, and data gaps.

After a pre-processing phase, we obtain initial smooth estimates of both diagonal and off-diagonal elements in the matrix $\mathbf{C}$. 
Subsequently, the \texttt{RJMCMC} algorithm is employed to update $S_{XX, \rm fit}$ and $S_{XY, \rm fit}$ alternately. Each update comprises 1000 iterations, with each iteration requiring approximately 0.15 seconds.
Given the simplicity of this example, only 2 circulations are performed within the \texttt{Gibbs} sampling framework.
The entire program can be completed in $\sim$10 minutes using a single CPU core (Intel Xeon Gold 6330).

\begin{figure}[htbp]
    \centering
    \includegraphics[width=0.8\linewidth]{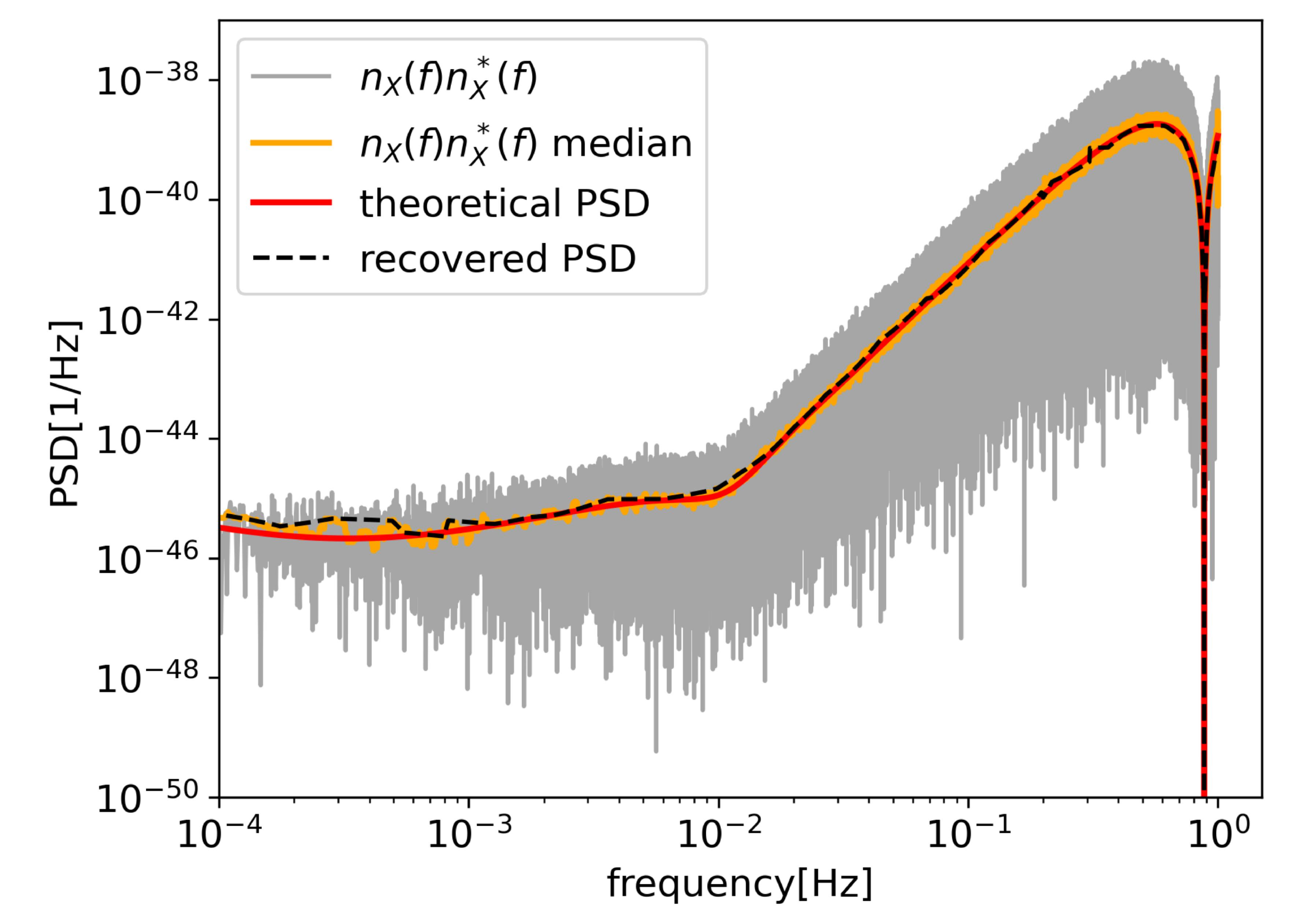}
    \caption{Recovered PSD (dashed black) compared with theoretical PSD (solid red) of $X$ channel on a logarithmic coordinate system. 
    The grey line shows the auto-correlated data spectrum, which is the noise in the $X$ channel multiplied by its complex conjugate. 
    The orange line is the median smoothed spectrum.}
    \label{fig:PSDfit}
\end{figure}

\begin{figure}[htbp]
    \centering
    \includegraphics[width=0.8\linewidth]{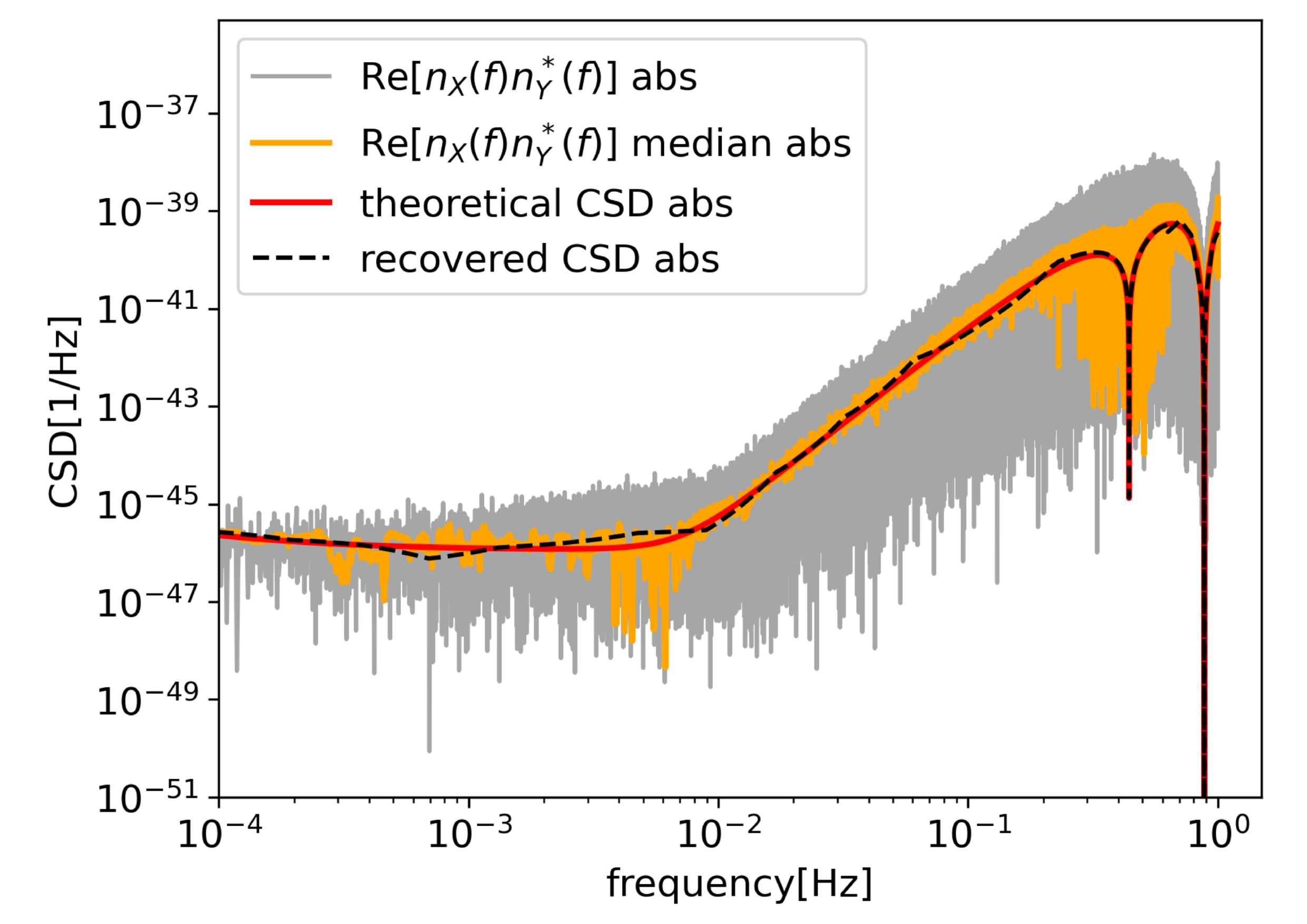}
    \includegraphics[width=0.8\linewidth]{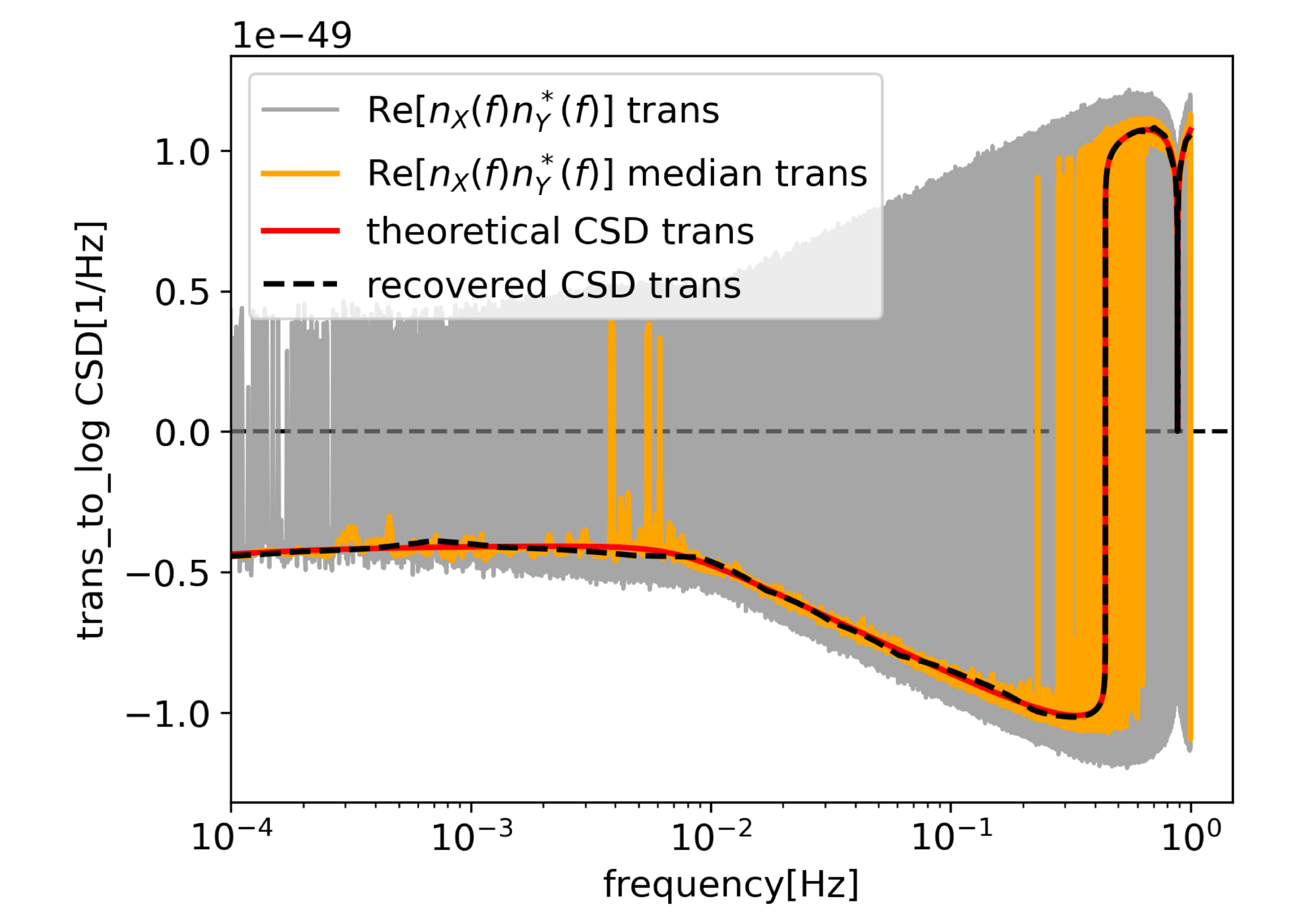}
    \caption{Recovered CSD (dashed black) compared with theoretical CSD (solid red) of $XY$ channel. The grey line shows the cross-correlated data spectrum, which is the noise in the $X$ channel multiplied by the complex conjugate of the noise in the $Y$ channel. 
    The orange line is the median smoothed spectrum. 
    The upper panel displays the result after taking the absolute value. 
    The bottom panel shows the log-transformed result in both positive and negative regions.}
    \label{fig:CSDfit}
\end{figure}

FIG.~\ref{fig:PSDfit} shows the PSD recovered by \texttt{NOISAR} (dashed black) compared to the theoretical PSD (solid red) of the $X$ channel, plotted in logarithmic coordinates. 
As shown in the figure, the proposed method successfully recovers the fluctuations of the PSD curve in the low-frequency region and the descending peak in the high-frequency region. 
FIG.~\ref{fig:CSDfit} shows the corresponding results for the CSD fitting. 
The upper panel shows the absolute values of the cross-correlated data spectrum and the CSD. 
To demonstrate the software performance in the negative region of the CSD, the bottom panel displays the logarithmic transformation of the data spectrum and the CSD curve, using the log method described in Sec. \ref{sec:log}. 
It can be seen that our method accurately captures the feature of the cross-correlated spectrum, which transitions from negative to positive values and descends back to zero. 

\begin{figure}[htbp]
    \centering
    \includegraphics[width=0.8\linewidth]{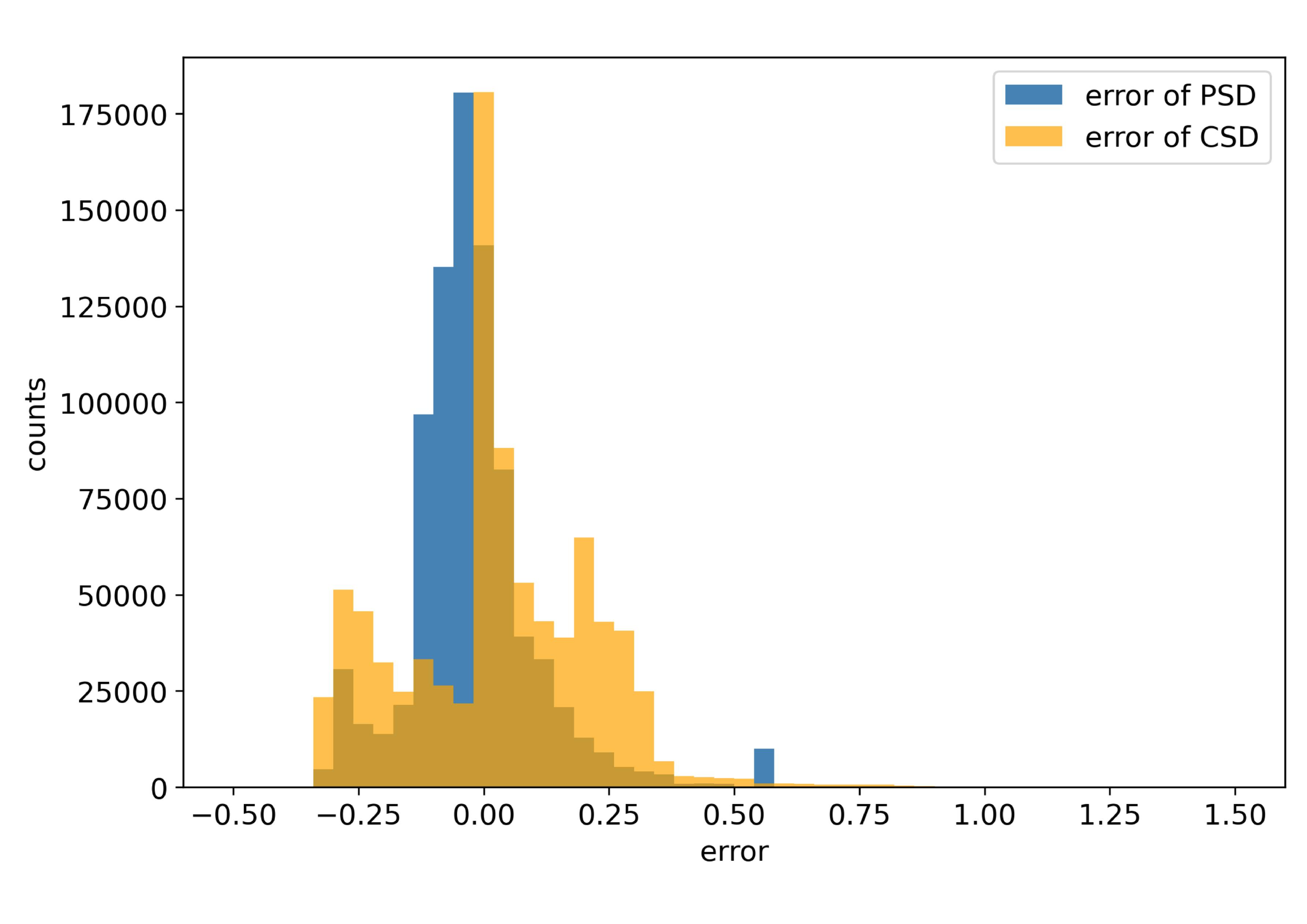}
    \caption{Histogram of the relative errors at each frequency bin. The blue part shows the result of the PSD error, and the orange part shows the result of the CSD error.}
    \label{fig:error}
\end{figure}

We calculate the relative error between the recovered spectrum $S_{\rm rec}$ and the theoretical one $S_{\rm theo}$, which is $(S_{\rm rec}-S_{\rm theo})/{S_{\rm theo}}$. 
FIG. \ref{fig:error} shows the histogram of pointwise errors at each frequency bin. Most of the errors are smaller than $20\%$. 
The PSD is estimated with a $9\%$ absolute error averaged over the frequency domain.
The CSD is estimated with a $15\%$ error.
In addition, we have applied the Bayesline method to fit the PSD of pure Gaussian noise for TianQin. The result is comparable to that of \texttt{NOISAR} in the low-frequency region. In the high-frequency region, \texttt{NOISAR} performs better at the zero point.

\section{Discussion and conclusion}
\label{sec:conclusion}

This paper advances noise modeling for multichannel systems by considering the full noise covariance matrix, including non-zero off-diagonal elements.
We develop the software \texttt{NOISAR} for adaptive modeling of noise covariance in
space-based gravitational wave detectors. 
A semi-analytical noise modeling approach is adopted. We use spline interpolation to fit the low-frequency portion of the spectrum and apply a trigonometric function to capture the high-frequency features that drop to zero or cross through zero. 
The initial estimated spectrum is obtained solely by smoothing the noise data spectrum. It is then refined using trans-dimensional Bayesian inference. 
The entire process is phenomenological and does not require any parametric analytical expressions of the noise model.

In this paper, we implement the preliminary version of \texttt{NOISAR} in a simplified scenario. The PSDs of the $X$, $Y$, and $Z$ channels are identical, and the CSDs between them are also the same. We consider pure Gaussian noise. 
Using \texttt{NOISAR}, we can effectively recover the features of both auto- and cross-correlated spectrum densities, with an overall relative error of about $10\%$ compared to the injected theoretical ones.

This method is flexible and does not rely on prior theoretical models, making it well-suited for potential extension to more complex scenarios. 
In the present work, we adopt the simplifying assumptions $S_{XX}=S_{YY}=S_{ZZ}$ and $S_{XY}=S_{YZ}=S_{XZ}$, thereby reducing the problem to fitting a single PSD and a single CSD curve. 
In a more general case where the noise in different channels is not identical, all nine elements of the noise covariance matrix (Eq.~(\ref{eq:cov})) will be treated as parameters to be inferred. The overall modeling framework remains unchanged. 
The dimension of the parameter space increases from 2 to 9, substantially increasing the computational cost. Besides, in this work, we consider the first-generation TDI, under which Eq.~(\ref{eq:hpsd}) and~(\ref{eq:hcsd}) can be directly used to recover the zero points in the high-frequency region. 
However, for second-generation TDI configurations, the sinusoidal terms in those equations require additional coefficient parameters to account for unequal arm lengths, introducing at least two more dimensions into the parameter space. 
Integrating this software into a global analysis framework, which typically involves hundreds of thousands of iterations, poses a computational challenge. Further efforts in computational acceleration are necessary to ensure the feasibility of large-scale applications.

Since the method is phenomenological, if the data contains some residual due to inaccurate parameter estimation and GW signal subtraction, it would be interpreted as part of the noise and incorporated into the noise covariance matrix. 
In data analysis pipelines where noise and signals are iteratively fitted (e.g., global fit), this may not be a problem. The large number of updates in the program will gradually refine the modeling of both noise and signals, implemented within the \texttt{Gibbs} sampling framework. 
Although the initial iterations may yield unsatisfactory estimates for signal parameters and noise characteristics, the outputs from each module iteratively enhance the performance of the others. This cyclic process of refinement eventually leads to accurate recovery of both the GW signal parameters and the noise.

For future work, we plan to extend the method to accommodate more complex and realistic scenarios, including non-stationary noise such as glitches and data gaps in the data stream. 
These improvements aim to enhance the generality and applicability of the software for space-based gravitational wave (GW) data analysis. The developed software module can be integrated into the global fit pipeline as the noise modeling module, providing a full noise covariance matrix that can be incorporated into the likelihood function for GW parameter estimation.

\section*{Data availability statement}

The data that support the findings of this study are openly available at the following URL/DOI: \url{https://github.com/TianQinSYSU/Noisar}.

\section*{Acknowledgments}

This work has been supported in part by 
the National Key Research and Development Program of China 
(No. 2023YFC2206700).
YMH is also supported by the Natural Science Foundation of China (
Grants  No.  12173104, and No. 12261131504), and the science research grants from the China Manned Space Project.

\section*{References}

\bibliographystyle{unsrt}
\bibliography{reference}

\end{document}